\documentclass[12pt]{article}
\usepackage{mathrsfs}
\usepackage{amsmath}
\usepackage{color}
\usepackage{psfrag}
\usepackage{graphicx}
\usepackage{subfigure}
\usepackage{cite}

\newcommand{\bea}{\begin{eqnarray}}
\newcommand{\eea}{\end{eqnarray}}
\newcommand{\be}{\begin{equation}}
\newcommand{\ee}{\end{equation}}
\newcommand{\vs}[1]{\vspace{#1 mm}}

\newcommand{\dsl}{\pa \kern-0.5em /}

\newcommand{\pa}{\partial}

\newcommand{\nn}{\nonumber\\}

\newcommand{\ba}{\begin{array}}
\newcommand{\ea}{\end{array}}
\newcommand{\bit}{\begin{itemize}}
\newcommand{\eit}{\end{itemize}}

\newcommand{\nd}{\textrm{d}}
\begin{document}
\topmargin 0mm
\oddsidemargin 0mm

\begin{flushright}

USTC-ICTS-18-15\\

\end{flushright}

\vspace{2mm}

\begin{center}

{\Large \bf Remark on the open string pair production enhancement}

\vs{10}

{\large Qiang Jia and J. X. Lu}

\vspace{4mm}

{\em
Interdisciplinary Center for Theoretical Study\\
 University of Science and Technology of China, Hefei, Anhui
 230026, China\\
 
}

\end{center}

\vs{10}

\begin{abstract}
	Recent studies by one of the present authors along with his collaborators in \cite{Lu:2009yx, Lu:2009au, Lu:2017tnm, Lu:2018suj} show that there exist the so-called open string pair production for a possible simplest system of two Dp branes, placed parallel at a separation and with each carrying different electric flux, in Type II superstring theories.  Further this pair production can be greatly enhanced when a magnetic flux, sharing no common field strength index with the electric one, is added, implying then $p \ge 3$. Given this, one may wonder if further enhancement can be achieved by adding more magnetic flux(es) in a similar fashion. In this paper, we explore this possibility.  It turns out that adding more such magnetic flux diminishes rather than enhances the pair production rate.  This actually implies that the largest enhancement occurs at  $p = 3$ when the same realistic electric and magnetic fluxes are applied for all $p \ge 3$.  Curiously one of D3 branes may be our own world and if so, the enhancement gives a possible opportunity to detect the pair production, therefore  to test the underlying string theories.  
	
\end{abstract}

\newpage
\section{Introduction}
	Applying a constant electric field to the vacuum of quantum electrodynamics (QED) gives rise to the so-called Schwinger pair production \cite{Schwinger:1951nm}.  It is natural to ask if an analogous process exists in various string theories. This was pursued in unoriented bosonic string and type I superstring a while ago in \cite{Bachas:1992bh, Porrati:1993qd}. We focus in this paper on Dp branes in the oriented Type II superstring theories for which the worldvolume electric and magnetic fluxes can be considered. 
	
	A Dp brane in Type II superstring theories is a non-perturbative Bogomol'ny-Prasad-Sommereld (BPS) solitonic vacuum-like object (for example, see \cite{Duff:1994an}), preserving one half of spacetime supersymmetry, and as such it is stable.  Its dynamics can also be described by an oriented perturbative open string with its two ends obeying the so-called Neumann  boundary conditions along the brane directions and the Dirichlet ones along the directions transverse to the brane  \cite{Polchinski:1995mt} when the string coupling $g_{s}$ is small.  This open string is charge-neutral in the sense that its two ends carry charge $+$ and $-$, respectively, with total zero net-charge. 	
	
      Just like the virtual electron/positron pair in QED vacuum, we have here the pair of virtual open string/anti open string, created from the Dp brane vacuum at certain instant, existing for a short period of time, then annihilating back to the vacuum.  An observer on the brane can only sense the ends, not the whole,  of these open strings,  as  charged or anti charged particles. So the pair of virtual open string/anti open string appears to the brane observer as two pairs of virtual charged/anti charged particles, each of which consists of two nearby ends of the string pair.  So the present quantum fluctuations are different from those in QED.   	

In a spirit to the Schwinger pair production\cite{Schwinger:1951nm}, one would  also expect to produce the charged particle/anti-charged particle or the open string pair if a constant worldvolume  electric field  is applied to an isolated Dp brane. However,  the stringy computations give a null result  due to the open string being charge-neutral and its two ends  experiencing the same electric field \cite{Bachas:1992bh, Lu:2017tnm}.  This is consistent with that a Dp carrying a constant electric field is a 1/2 BPS non-threshold bound state \cite{Lu:1999uca}, therefore being stable rather than unstable. So this system cannot decay via the open string pair production. This can also be understood as the lack of the force on either of the open string or the anti open string in the pair to pull them apart since the net-force acting on either of them vanishes under the action of the constant applied electric flux less than its critical value\footnote{\label{footnote1}When the applied electric flux reaches its critical value unity, the forces, equal in magnitude but opposite in direction, acting at the two ends of the open string or anti open string can break it and the pair production instability develops. However, in practice, we never reach this critical electric flux in laboratory and so this will not be an issue for the purpose of the present discussion.}. In other words, there is no Schwinger-type pair production here.  
	
In order to have the open string pair production, a possibility is to let the two ends of the open string  experience different electric fields since the charge-neutral property of the string 
cannot be altered.  A possible simplest setup for this is to consider a system of two Dp branes placed parallel at a separation and with each carrying a different electric flux. The open strings produced, if any,  should come from those connecting the two Dp branes along the transverse directions, therefore related to the extra dimensions, from the viewpoint of the brane observer.   Stringy computations do give a non-vanishing pair production rate \cite{Lu:2009yx} but it is usually vanishingly small for any realistic electric flux applied, which is usually much smaller than the critical one.  This rate can however be greatly enhanced if we add also a magnetic flux in a way that the two fluxes do not share a common field strength index \cite{Lu:2009au, Lu:2017tnm, Lu:2018suj}. 

    We denote now the general worldvolume dimensionless flux on one Dp by $\hat F$ and on the other by $\hat F'$,  both being antisymmetric $(1 + p) \times (1 + p)$-matrices with the same structure. For this enhancement, we can choose, without loss of generality, the non-vanishing components for $\hat F$ as
\be\label{fstructure}
 \hat F_{01} = - \hat F_{10} = - \hat f_1,  \qquad \hat F_{23} = - \hat F_{32} = - \hat g_{2},
 \ee
  and the same non-vanishing components for $\hat F'$ but denoting each with a prime. Here $\hat f_1$ stands for the electric flux with $|\hat f_1| < 1$ and $\hat g_{2}$ the magnetic flux with $|\hat g_{2}| < \infty$. 
Note that the dimensionless flux $\hat F$ or $\hat F'$ is related to the corresponding laboratory one $F$ or $F'$ via $\hat F = 2\pi \alpha' F$ and $\hat F' = 2\pi \alpha' F'$ with $\alpha'$ the Regge slope parameter, related to the fundamental string tension $T_{f}$ as $T_{f}  = 1/(2\pi \alpha')$.  This structure of the fluxes concerning their non-vanishing components must imply $p \ge 3$.  

   The study given in \cite{Lu:2018suj} for various cases of two fluxes on each brane indicates that the most efficient and direct enhancement of the pair production is by adding a magnetic flux on each brane as specified above. This structure makes us wonder if further enhancement can be possible\footnote{\label{footnote2} In general, placing an infinitely extended Dp in spacetime will cause it to curve. For our purpose,  we try to avoid this to happen at least to the probe distance in which we are interested. For this, we need to limit our discussion in this paper to $p \le 6$ cases since these Dp branes have well-behaved supergravity configurations which are all asymptotically flat. Moreover, when the string coupling is small, i.e. $g_{s} \ll 1$, placing one such Dp in spacetime will keep the spacetime flat even for a probe distance to the brane in the substringy scale $\alpha'^{1/2} \gg r \gg g^{1/(7 - p)}_{s} \alpha'^{1/2}$   as discussed in section 2 of \cite{Lu:2007kv}. So this indeed meets our requirement. }. For example, for $p \ge 5$, we might have a further enhancement of the rate if we add one more magnetic flux, in addition to the above (\ref{fstructure}), of the form $\hat F_{45} = - \hat F_{54} = - \hat g_{3}$ (and in a similar fashion for $\hat F'$). As discussed in footnote \ref{footnote2}, this is actually the only  case possible for this purpose.  Either answer of this investigation will be important for us to determine which system can give rise to the largest possible pair production.  
       
   In this paper, we investigate this possible enhancement.  It turns out that adding such additional magnetic flux diminishes rather than enhances the pair production, assuming the fluxes given in (\ref{fstructure}) remain the same. This result  implies that the pair production rate computed using fluxes in (\ref{fstructure}), in the absence of the additional fluxes $\hat g_{3}$ and $\hat g'_{3}$,  is the largest for each $p \ge 5$. Combined this with the previous result given in \cite{Lu:2017tnm},  we conclude that the rate for the system of two D3 branes is the largest among the $p \ge 3$  cases when we take the same fluxes as given in (\ref{fstructure}) for all the cases.   Curiously this is interesting since one of the D3 can be our own 4-dimensional world.  This also makes it a possibility for its detection.
   
     As  mentioned earlier, the pair production is related to the open strings connecting the two Dp branes, therefore to the dimensions transverse to the branes. From the perspective of the brane observer, the open string pair appears as the charged particle/anti-charged particle pair and the transverse dimensions appear as extra-dimensions.  So a detection of this pair production by the brane observer implies the existence of extra dimensions.  It also provides a test of the underlying string theories.  
   
   This paper is organized as follows.  In section 2, we follow the steps outlined in \cite{Lu:2018suj} using the closed string boundary states to present first the closed string cylinder interaction amplitude in terms of various $\theta$-functions between the two Dp branes with one additional magnetic flux on each brane for $p \ge 5$. We will specify the structure of these fluxes in this section explicitly.  We then use Jacobi transformation to pass this amplitude to the open string annulus one for each case.  In section 3, we will use this open string annulus amplitude to obtain the non-perturbative open string pair production rate following \cite{Porrati:1993qd, Lu:2009au, Lu:2017tnm}, analyze this rate and discuss its significance and implications. We discuss and conclude this paper in section 4. 
   
\section{The open string annulus amplitude} 
In this section, we will obtain and study the open string annulus interaction amplitude between two Dp-branes, as specified earlier,  with one additional magnetic flux on each brane for $p\ge 5$.  The simpler approach, when both electric and magnetic fluxes are present, is to compute the corresponding closed string cylinder amplitude first.  This is due to that each such Dp brane carrying these fluxes can be easily represented by its closed string boundary state (for example, see \cite{Di Vecchia:1999rh, DiVecchia:1999uf}).  The cylinder amplitude can be easily obtained following the steps outlined in \cite{Lu:2018suj, Di Vecchia:1999rh}.  
Once having this amplitude, we can obtain the corresponding open string annulus one simply by a Jacobi transformation on this one. 

\subsection{The closed string cylinder amplitude}

Following \cite{Lu:2018suj}, we now briefly outline the steps to compute the closed string cylinder interaction amplitude between two Dp, placed parallel at a separation and with one brane carrying the  fluxes 
\begin{equation}\label{F}
	\hat{F} = \left[ \begin{array}{cccccccc}
	0&- \hat f_1&0&0&0&0&0&\cdots\\
	\hat f_1&0&0&0&0&0&0&\cdots\\
	0&0&0&- \hat g_2&0&0&0&\cdots\\
	0&0&\hat g_2&0&0&0&0&\cdots\\
	0&0&0&0&0&- \hat g_3&0&\cdots\\
	0&0&0&0&\hat g_3&0&0&\cdots\\
	0&0&0&0&0&0&0&\cdots\\
	\vdots&\vdots&\vdots&\vdots&\vdots&\vdots&\vdots&\ddots
	\end{array} \right]_{(p+1)\times(p+1)},
\end{equation}
and the other carrying the fluxes 
\begin{equation}\label{F'}
	\hat{F'} = \left[ \begin{array}{cccccccc}
	0&- \hat f'_1&0&0&0&0&0&\cdots\\
	\hat f'_1&0&0&0&0&0&0&\cdots\\
	0&0&0&- \hat g'_2&0&0&0&\cdots\\
	0&0&\hat g'_2&0&0&0&0&\cdots\\
	0&0&0&0&0&- \hat g'_3&0&\cdots\\
	0&0&0&0&\hat g'_3&0&0&\cdots\\
	0&0&0&0&0&0&0&\cdots\\
	\vdots&\vdots&\vdots&\vdots&\vdots&\vdots&\vdots&\ddots
		\end{array} \right]_{(p+1)\times(p+1)}.
\end{equation}
From the non-vanishing components of $\hat F$ or $\hat F'$, it is clear $p \ge 5$.  There are in general two contributions to this amplitude, one from the so-called NS-NS sector and the other from the R-R sector, respectively.  The total closed string cylinder amplitude is the sum of the two after the so-called Gliozzi-Scherk-Olive (GSO) projection for each, namely, 
\be\label{totalamplit}
\Gamma = \Gamma_{\textrm{NSNS}} + \Gamma_{\textrm{RR}}, 
\ee
where the GSO projected amplitudes in both NSNS-sector and RR-sector are 
\be\label{ns-r-amplitude}
\Gamma_{\textrm{NSNS}} =  \frac{1}{2}\left[\Gamma_{\textrm{NSNS}} (+) - \Gamma_{\textrm{NSNS}} (-) \right] , \quad  \Gamma_{\textrm{RR}} = \frac{1}{2}\left[\Gamma_{\textrm{RR}} (+) + \Gamma_{\textrm{RR}} (-) \right].
\ee
Here  $\Gamma_{\rm NSNS} (\pm)$  and $\Gamma_{\rm RR} (\pm)$ can be read, for the present case, from the corresponding general amplitudes, respectively, given in \cite{Lu:2018suj}.  We have then, with $\eta \eta' = \pm$, 
\bea\label{amplitudensns}
\Gamma_{\rm NSNS} (\eta'\eta) &=& \frac{V_{p + 1} \left[(1 - \hat f^{2}_{1})(1 - \hat f'^{2}_{1})(1 + \hat g^{2}_{2})(1 + \hat g'^{2}_{2})(1 + \hat g^{2}_{3})(1 + \hat g'^{2}_{3})\right]^{\frac{1}{2}}}{(8 \pi^2 \alpha')^{\frac{1 + p}{2}}} \int_0^\infty \frac{d t} {t^{\frac{9 - p}{2}}} \frac{e^{- \frac{y^2}{2\pi\alpha' t}}}{ |z|}\nn
&\,&\times\prod_{n = 1}^\infty \left(\frac{1  + \eta'\eta |z|^{2n - 1}}{1 - |z|^{2n}}\right)^{2} \prod_{a = 1}^3 \frac{(1 + \lambda_a \eta'\eta |z|^{2 n - 1})(1 + \lambda^{-1}_a \eta'\eta |z|^{2 n - 1})}{ (1 - \lambda_a |z|^{2n})(1 - \lambda^{-1}_a |z|^{2n})},
\eea 
and 
\bea\label{amplituderr}
\Gamma_{\rm RR} (\eta'\eta) &=& \frac{V_{p + 1} \left[(1 - \hat f^{2}_{1})(1 - \hat f'^{2}_{1})(1 + \hat g^{2}_{2})(1 + \hat g'^{2}_{2})(1 + \hat g^{2}_{3})(1 + \hat g'^{2}_{3})\right]^{\frac{1}{2}}}{(8\pi^2\alpha')^{\frac{1 + p}{2}}}  {}_{\rm 0R}\langle B', \eta'| B, \eta\rangle_{\rm 0R}\nn
&\,&\times \int_0^\infty \frac{dt}{t^{\frac{9 - p}{2}}} \, e^{- \frac{y^2}{2\pi \alpha' t}}  \prod_{n = 1}^\infty \left(\frac{1 + \eta' \eta |z|^{2n}}{1 - |z|^{2n}}\right)^{2} \prod_{a = 1}^3 \frac{(1 + \eta'\eta \lambda_a |z|^{2n})(1 + \eta'\eta \lambda^{-1}_a |z|^{2n})}{(1 - \lambda_a |z|^{2n})(1 - \lambda^{-1}_a |z|^{2n})}.\nn
\eea
In the above, $V_{p + 1}$ denotes the volume of the brane worldvolume, $|z| = e^{- \pi t} < 1$, $y$ the brane separation and the zero-mode contribution in the RR-sector can be evaluated with the fluxes given in (\ref{F}) and (\ref{F'}), following \cite{Yost, Billo:1998vr}, as 
\bea\label{0mme}
{}_{\rm 0R}\langle B', \eta'| B, \eta\rangle_{\rm 0R} &\equiv& {}_{\rm 0R}\langle B'_{\rm sgh}, \eta'| B_{\rm sgh}, \eta\rangle_{\rm 0R} \times {}_{\rm 0R}\langle B'_\psi, \eta'| B_\psi, \eta\rangle_{\rm 0R} ,\nn
&=& - \frac{2^4 (1- \hat f_1 \hat f'_1)(1+ \hat g_2 \hat g'_2)(1+ \hat g_3 \hat g'_3)}{\sqrt{(1-\hat f^{2}_1) (1- \hat f'^{2}_1)(1+ \hat g^{2}_2) (1+\hat g'^{2}_2)(1+\hat g^{2}_3) (1+ \hat g'^{2}_3)}} \delta_{\eta \eta' , +}.\,
\eea
where $| B_{\rm sgh}, \eta\rangle$ denotes the boundary state of superghosts $\beta$ and $\gamma$ while $| B_\psi, \eta\rangle$ the boundary state of matter field $\psi^{\mu}$. 
The $\lambda_a$ for $a = 1, 2, 3$ are given, respectively, as
\bea\label{lambda}
	\lambda_1 + \lambda_1^{-1} &=& 2\frac{(1+\hat f_1^2)(1 + \hat f'^{2}_1)-4 \hat f_1 \hat f'_1}{(1-\hat f^{2}_1) (1- \hat f'^{2}_1)},\nn
	\lambda_2 + \lambda_2^{-1} &=& 2\frac{(1- \hat g_2^2)(1- \hat g'^{2}_2)+4 \hat g_2 \hat g'_2}{(1+ \hat g^{2}_2) (1+ \hat g'^{2}_2)},\nn
	\lambda_3 + \lambda_3^{-1} &=& 2\frac{(1- \hat g_3^2)(1- \hat g'^{2}_3)+4 \hat g_3 \hat g'_3}{(1+ \hat g^{2}_3) (1+ \hat g'^{2}_3)}.
\eea 
We have then the total amplitude from (\ref{amplituderr}), (\ref{amplitudensns}) and  (\ref{totalamplit}) as
\bea\label{totala}
\Gamma &=& \frac{V_{p + 1}\left[(1-\hat f^{2}_1) (1- \hat f'^{2}_1)(1+ \hat g^{2}_2) (1+ \hat g'^{2}_2)(1+ \hat g^{2}_3) (1+ \hat g'^{2}_3)\right]^{\frac{1}{2}} }{2 (8 \pi^2 \alpha')^{\frac{1+p}{2}}}\int_0^{\infty} \frac{\nd t}{t^{\frac{9-p}{2}}} e^{-\frac{y^2}{2\pi \alpha' t}} \nn
	&\,& \times \left[|z|^{-1}\left( \prod_{n=1}^{\infty} A_n - \prod_{n=1}^{\infty} B_n \right) - 2^4 \cos \pi \nu \cos \pi \nu' \cos \pi \nu'' \prod_{n=1}^{\infty} C_n   \right],
\eea
where 
\begin{equation}
	A_n = \left(\frac{1+|z|^{2n-1}}{1-|z|^{2n}} \right)^2 \prod_{a=1}^3 \frac{(1+\lambda_{a} |z|^{2n-1})(1+\lambda_a^{-1} |z|^{2n-1})}{(1-\lambda_{a} |z|^{2n})(1-\lambda_a^{-1} |z|^{2n})}, \nonumber
\end{equation}
\begin{equation}
	B_n = \left(\frac{1-|z|^{2n-1}}{1-|z|^{2n}} \right)^2 \prod_{a=1}^3 \frac{(1-\lambda_{a} |z|^{2n-1})(1-\lambda_a^{-1} |z|^{2n-1})}{(1-\lambda_{a} |z|^{2n})(1-\lambda_a^{-1} |z|^{2n})}, \nonumber
\end{equation}
\begin{equation}
	C_n = \left(\frac{1+|z|^{2n}}{1-|z|^{2n}} \right)^2 \prod_{a=1}^3 \frac{(1+\lambda_{a} |z|^{2n})(1+\lambda_a^{-1} |z|^{2n})}{(1-\lambda_{a} |z|^{2n})(1-\lambda_a^{-1} |z|^{2n})}.
\end{equation}
In the above, we have set $\lambda_1 = e^{2\pi i \nu},\lambda_2 = e^{2\pi i \nu'}$ and $\lambda_3= e^{2\pi i \nu''}$, and we also make use of (\ref{0mme}) and (\ref{lambda}).  We now express the amplitude in terms of various $\theta$-functions and the Dedekind $\eta$-function, following their standard definitions given, for example, in  \cite{polbookone}. We have then the amplitude (\ref{totala}) as 
\bea\label{tamplitude-theta}
\Gamma &=& \frac{ 2^{2}\, i\, V_{p + 1} |\hat f_{1} - \hat f'_{1}|\, |\hat g_{2} - \hat g'_{2}|\, |\hat g_{3} - \hat g'_{3}|}{(8 \pi^2 \alpha')^{\frac{1+p}{2}}}\int_0^{\infty} \frac{\nd t}{t^{\frac{9-p}{2}}} \frac{e^{-\frac{y^2}{2\pi \alpha' t}} }{\eta^{3} (it)\, \theta_{1} (\nu | it) \theta_{1} (\nu' |it) \theta_{1} (\nu'' | it) } \nn
&\,& \times \left[  \theta_{3} (0 |it) \theta_{3} (\nu | it) \theta_{3} (\nu' |it) \theta_{3} (\nu'' | it) -  \theta_{4} (0 |it) \theta_{4} (\nu | it) \theta_{4} (\nu' |it) \theta_{4} (\nu'' | it) \right.\nn
&\,& \,\, \left.- \theta_{2} (0 |it) \theta_{2} (\nu | it) \theta_{2} (\nu' |it) \theta_{2} (\nu'' | it)\right],\nn
&=&  \frac{ 2^{3}\, i\, V_{p + 1} |f_{1} - f'_{1}|\, |g_{2} - g'_{2}|\, |g_{3} - g'_{3}|}{(8 \pi^2 \alpha')^{\frac{1+p}{2}}}\int_0^{\infty} \frac{\nd t}{t^{\frac{9-p}{2}}}\frac{ e^{-\frac{y^2}{2\pi \alpha' t}} }{\eta^{3} (it)}\nn
&\,&\times \frac{\theta_{1} \left(\left.\frac{\nu +\nu' + \nu''}{2} \right|  it \right) \theta_{1} \left(\left.\frac{ \nu - \nu' - \nu''}{2} \right| it \right) \theta_{1} \left(\left.\frac{\nu - \nu' + \nu''}{2} \right| it \right)  \theta_{1} \left(\left.\frac{\nu +\nu' - \nu''}{2} \right| it \right)} { \theta_{1} (\nu | it) \theta_{1} (\nu' |it) \theta_{1} (\nu'' | it) },
\eea
where in the last equality we have made use of the identity 
\bea\label{identity}
&&2 \,\theta_{1} \left(\left.\frac{\nu +\nu' + \nu''}{2}\right| it \right) \theta_{1} \left(\left.\frac{\nu - \nu' - \nu''}{2} \right| it \right) \theta_{1} \left(\left.\frac{\nu - \nu' + \nu''}{2}\right| it \right)  \theta_{1} \left(\left.\frac{\nu +\nu' - \nu''}{2}\right| it \right) \nn
&&= \left[  \theta_{3} (0 |it) \theta_{3} (\nu | it) \theta_{3} (\nu' |it) \theta_{3} (\nu'' | it) -  \theta_{4} (0 |it) \theta_{4} (\nu | it) \theta_{4} (\nu' |it) \theta_{4} (\nu'' | it) \right.\nn
&& \qquad \left.- \theta_{2} (0 |it) \theta_{2} (\nu | it) \theta_{2} (\nu' |it) \theta_{2} (\nu'' | it)\right],
\eea
which is a special case of the more general identity given in \cite{whittaker-watson}.  Given (\ref{lambda}),  the parameter $\nu$ is actually imaginary and can be set as $\nu = i \nu_{0}$ with 
$\nu_{0} \in [0, \infty)$ while both $\nu'$ and $\nu''$ are real and for convenience can be set as $\nu' = \nu_{1}$ and $\nu'' = \nu_{2}$ with $\nu_{1}, \nu_{2} \in [0, 1]$.  In terms of the applied fluxes, we have from (\ref{lambda})
\bea\label{nu0nu1nu2}
	\cosh \pi \nu_0 &=& \frac{1- \hat f_1 \hat f'_1}{\sqrt{(1-\hat f^2_1)(1-\hat f'^{2}_1)}}, \quad \sinh \pi \nu_0 = \frac{|\hat f_1 - \hat f'_1|}{\sqrt{(1-\hat f^2_1)(1-\hat f'^{2}_1)}};\nn
	\cos \pi \nu_1 &=& \frac{1+\hat g_2 \hat g'_2}{\sqrt{(1+ \hat g^2_2)(1+\hat g'^{2}_2)}},\quad \sin \pi \nu_1 = \frac{|\hat g_2 - \hat g'_2|}{\sqrt{(1+ \hat g^2_2)(1+\hat g'^{2}_2)}};\nn
	\cos \pi \nu_2 &=& \frac{1+\hat g_3 \hat g'_3}{\sqrt{(1+ \hat g^2_3)(1+\hat g'^{2}_3)}},\quad \sin \pi \nu_2 = \frac{|\hat g_3 - \hat g'_3|}{\sqrt{(1+\hat g^2_3)(1+\hat g'^{2}_3)}}.
\eea
The last equality of the amplitude (\ref{tamplitude-theta}) can be expressed explicitly as 
\bea\label{tae}
\Gamma &=& \frac{ 4\, K\, V_{p + 1} [\cosh \pi \nu_{0} - \cos\pi(\nu_{1} + \nu_{2})][\cosh\pi\nu_{0} - \cos\pi(\nu_{1} - \nu_{2})] }{(8 \pi^2 \alpha')^{\frac{1+p}{2}}}\int_0^{\infty} \frac{\nd t}{t^{\frac{9-p}{2}}} e^{-\frac{y^2}{2\pi \alpha' t}}\nn
&\,&\times \prod_{n = 1}^{\infty} D_{n},
\eea
where
\be
	K = \left[(1-\hat f^2_1)(1-\hat f'^2_1)(1+\hat g^2_2)(1+\hat g'^2_{2})(1+\hat g^2_3)(1+\hat g'^2_{3})\right]^{\frac{1}{2}},
\ee
and 
\be
D_n = \frac{\prod_{a=0}^{1} \prod_{b=0}^{1} \left[1-2|z|^{2n}e^{(-)^a \pi \nu_0} \cos \pi (\nu_1+(-)^b \nu_2) + |z|^{4n} e^{(-)^a 2\pi \nu_0}\right]}{(1-|z|^{2n})^2 (1-2|z|^{2n}\cosh 2\pi \nu_0 + |z|^{4n}) \prod_{i=1}^{2}(1-2|z|^{2n}\cos 2\pi \nu_i + |z|^{4n}) }.
\ee
For large brane separation $y$, the main contribution to the amplitude (\ref{tae}) comes from the large $t$ integration for which $|z| = e^{-\pi t} \to 0$ and  $D_{n} \approx 1$.  It is then clear that $\Gamma > 0$, giving an attractive interaction by our conventions between the two Dp branes as expected.  For small $y$,  the small $t$ contributes also to the amplitude. Note that in $D_{n}$ both $1-2|z|^{2n}e^{\pm \pi \nu_0} \cos \pi (\nu_1 \pm \nu_2) + |z|^{4n} e^{\pm 2\pi \nu_0} \ge (1-|z|^{2 n} e^{\pm \pi \nu_0})^2 > 0$  and $(1-2|z|^{2n}\cos 2\pi \nu_i + |z|^{4n}) \ge (1-|z|^{2 n})^2 > 0$ but the factor $1-2|z|^{2n}\cosh 2\pi \nu_0 + |z|^{4n}$ in the denominator can be negative for small enough $t$. So this gives an ambiguity about the nature of the interaction for small $y$ since there are an infinite number of such factors appearing in infinite product in the amplitude.  So we expect some interesting physics to occur for small $y$. As will be seen, this can be more transparent in the open string channel which we turn next. 

\subsection{The open string annulus amplitude}
In passing the last equality of the closed string cylinder amplitude in (\ref{tamplitude-theta}) to the open string annulus one, we need to use the Jacobi transformation $t\rightarrow t' = \frac{1}{t}$. For this, we also need the following relations for the $\theta_{1}$-function and the Dedekind $\eta$-function,
\be\label{jacobi}
\eta (\tau) = \frac{1}{(- i \tau)^{1/2}} \eta \left(- \frac{1}{\tau}\right), \quad \theta_1 (\nu | \tau) = i \frac{e^{- i \pi \nu^2/\tau}}{(- i \tau)^{1/2}} \theta_1 \left(\left.\frac{\nu}{\tau} \right| - \frac{1}{\tau}\right).
\ee 
The open string annulus amplitude is then   
\bea\label{annulusa}
\Gamma &=&-\frac{2^3  V_{p+1} |\hat f_1- \hat f'_1| \,|\hat g_2- \hat g'_2|\,|\hat g_3- \hat g'_3|}{(8\pi^2 \alpha')^{\frac{p+1}{2}}} \int_0^{\infty} \frac{\nd t'}{t'^{\frac{p-3}{2}}}e^{-\frac{y^2 t'}{2\pi \alpha'}} \nn
&& \times \frac{\theta_1\left(\left.\frac{\nu_0-i(\nu_1+\nu_2)}{2}t'\right| it' \right)\theta_1\left(\left.\frac{\nu_0+i(\nu_1+\nu_2)}{2}t'\right|it'\right)\theta_1\left(\left.\frac{\nu_0-i(\nu_1-\nu_2)}{2}t'\right|it'\right)\theta_1\left(\left.\frac{\nu_0+i(\nu_1-\nu_2)}{2}t'\right|it'\right)}{\eta^3(it') \theta_1(\nu_0 t'|it') \theta_1(i \nu_1 t'|it') \theta_1(i \nu_2 t'|it')},\nn
&=& \frac{4\, V_{p + 1}\, |\hat f_1 - \hat f'_1|\,|\hat g_2 - \hat g'_2|\,|\hat g_3 - \hat g'_3|}{(8\pi^2 \alpha')^{{\frac{p+1}{2}}}} \int_0^{\infty} \frac{\nd t}{t^{\frac{p-3}{2}}} e^{-\frac{y^2 t}{2 \pi \alpha'}} E(\nu_{0}, \nu_{1}, \nu_{2}; t)  \prod_{n=1}^{\infty} Z_n,
\eea
where in the last equality we have dropped the prime on $t$, 
\be\label{enhf}
E (\nu_{0}, \nu_{1}, \nu_{2}; t) = \frac{\left[\cosh\pi(\nu_{1} + \nu_{2}) t - \cos\pi\nu_{0} t \right]\left[\cosh\pi(\nu_{1} - \nu_{2}) t - \cos\pi\nu_{0} t\right]}{\sin\pi\nu_{0} t \sinh\pi \nu_{1} t \sinh\pi\nu_{2} t},
\ee
and
\be\label{zf}
Z_n =  \frac{\prod^{1}_{a=0} \prod^{1}_{b=0}\left[1 - 2 |z|^{2n} e^{(-)^a \pi \left(\nu_{1} + (-)^{b} \nu_{2}\right) t} \cos \pi \nu_{0} t + |z|^{4n} e^{(-)^a 2\pi \left(\nu_{1} + (-)^{b}\nu_{2}\right)t }\right]  }{(1-|z|^{2n})^2 (1-2|z|^{2n} \cos 2\pi \nu_0 t + |z|^{4n})\prod^{2}_{i=1}(1-2|z|^{2n} \cosh 2\pi \nu_i t + |z|^{4n})}.
\ee
Note that $Z_{n} > 0$ since $n \ge 1$, $|z| = e^{- \pi t}$ and $\nu_{1}, \nu_{2} \in [0, 1]$. The interesting physics comes from the  factor $E (\nu_{0}, \nu_{1}, \nu_{2}; t)$ given in (\ref{enhf}). In the integral representation of the amplitude (\ref{annulusa}), we actually have $t > 0$.  $E (\nu_{0}, \nu_{1}, \nu_{2}; t)$ blows up at points $t_{k} = k/\nu_{0}$ with $k = 1, 2 \cdots$, which are determined from the zeros of the factor $\sin \pi \nu_{0} t$ in its denominator.  In other words, the integrand has an infinite number of simple poles along the 
positive t-axis.  This indicates that the amplitude has an imaginary part which gives rise to the decay of the underlying system. As we will turn to the next section, this decay is to release the excess energy due to the applied fluxes, via the so-called open string pair production, to relax the underlying system to become a new BPS one. 

\section{The open string pair production enhancement}
Given what has been said in the previous section, we now compute the non-perturbative decay or the so-called pair production rate from the open string annulus amplitude given in the second equality of (\ref{annulusa}).  Following \cite{Bachas:1992bh}, this rate of pair production per unit worldvolume can be obtained as the
imaginary part of this amplitude and is given as the sum of the residues of the poles of the
 integrand  times $\pi$,
\bea\label{pprate}
{\cal W} &=& - \frac{2\, {\rm Im} \Gamma}{V_{p+1}},  \nn
	&=& \frac{8 |\hat f_1 - \hat f'_1||\hat g_2 - \hat g'_2||\hat g_3 - \hat g'_3|}{(8 \pi^2 \alpha')^{{\frac{p+1}{2}}}} \sum_{k=1}^{\infty} (-)^{k-1} \left(\frac{\nu_0}{k}\right)^{\frac{p-5}{2}}  e^{-\frac{k y^2}{2\pi \alpha' \nu_{0}}} \, E_{k}\, \prod_{n=1}^{\infty} Z_{n, k},	
\eea
where
\be\label{ek}
E_{k} = \frac{\left[\cosh\frac{\pi (\nu_{1} + \nu_{2}) k}{\nu_{0}} - (-)^{k}\right]\, \left[\cosh\frac{\pi (\nu_{1} - \nu_{2}) k}{\nu_{0}} - (-)^{k}\right]}{k \sinh\frac{\pi \nu_{1} k}{\nu_{0}} \sinh\frac{\pi \nu_{2} k}{\nu_{0}}},
\ee
and
\be
Z_{n,k} = \left(1 - e^{- \frac{2\pi n k}{\nu_{0}}}\right)^{- 4}\prod^{1}_{a= 0}\frac{\left[1-(-)^k e^{-\frac{2\pi  k}{\nu_0}\left(n- (-)^{a}\frac{\nu_1 + \nu_2}{2}\right)}\right]^2 \left[1-(-)^k e^{-\frac{2\pi  k}{\nu_0}\left(n- (-)^{a}\frac{\nu_1 - \nu_2}{2}\right)}\right]^2 }{\left[1-e^{-\frac{2k\pi}{\nu_0}\left(n- (-)^{a}\nu_1\right)}\right] \left[1-e^{-\frac{2k\pi}{\nu_0}\left(n- (-)^{a}\nu_2\right)}\right]}.
\ee
Recall here $p \ge 5$.   As in the previous studies \cite{Lu:2009au, Lu:2017tnm, Lu:2018suj}, for given fluxes, the larger the brane separation $y$ is and/or the larger the integer $k$ is, the smaller the rate is since the corresponding open string, whose mass is given by $k y T_{f}$ with $T_{f} = 1/(2\pi \alpha')$ the fundamental string tension, is more massive, therefore more difficult to be produced.  The larger $|\hat f_{1} - \hat f'_{1}|$ is, therefore the larger $\nu_{0}$, the larger the rate is.  When either of $\hat f_{1}$ or $\hat f'_{1}$ reaches  its critical value unity, we have $\nu_{0} = \infty$ from the first two equations in (\ref{nu0nu1nu2}).  Then $Z_{n, k}$ blows up for odd $k$ and so does the rate, the occurring of the pair production instability as mentioned earlier in footnote \ref{footnote1}.  This rate becomes the one given in \cite{Lu:2018suj} when we set $\hat g_{3} = \hat g'_{3} = 0$ for which $\nu_{2} = 0$ from (\ref{nu0nu1nu2}).  

For realistic laboratory electric fluxes, we expect in general $|\hat f_{1}|, |\hat f'_{1}| \ll 1$ and $|\hat f_{1} - \hat f'_{1}| \ll 1$. So we have $\nu_{0} \ll 1$ from (\ref{nu0nu1nu2}) which implies $Z_{n, k} \approx 1$ and the rate can be approximated by the leading $k = 1$ term.  We have also $|\hat g_{2}|, |\hat g_{3}|, |\hat g'_{2}|, |\hat g'_{3}| \ll 1$. So $\nu_{1}$ and $\nu_{2}$ are both small and we can replace both $|\hat g_{2} - \hat g'_{2}|$ and $|\hat g_{3} - \hat g'_{3}|$  by their respective $\pi \nu_{1}$ and $\pi \nu_{2}$ from (\ref{nu0nu1nu2}). So we have the rate (\ref{pprate}) as
\be\label{approxrate}
{\cal W} \approx \frac{8 \pi^{2} |\hat f_1 - \hat f'_1| \nu_{1} \nu_{2}}{(8 \pi^2 \alpha')^{{\frac{p+1}{2}}}}  \nu_0^{\frac{p-5}{2}}  e^{-\frac{ y^2}{2\pi \alpha' \nu_{0}}}\,  \frac{\left[\cosh\frac{\pi (\nu_{1} + \nu_{2}) }{\nu_{0}} +1 \right]\left[\cosh\frac{\pi (\nu_{1} - \nu_{2})}{\nu_{0}} + 1\right]}{\sinh\frac{\pi \nu_{1} }{\nu_{0}} \sinh\frac{\pi \nu_{2} }{\nu_{0}}}.
\ee 
Note that the above rate is symmetric to $\nu_{1}$ and $\nu_{2}$ as it should be. So without loss of generality, we assume  $\nu_{1} \ge \nu_{2}$  from now on. This is  consistent with the actual practice in that we always want to obtain the largest enhancement before turning on the second set of magnetic fluxes $\hat g_{3}$ and $\hat g'_{3}$. This also gives us the option to set $\hat g_{3} = \hat g'_{3} = 0$ (so $\nu_{2}= 0$) in the above rate.  
With these, we are now ready to compare this rate (\ref{approxrate}) with $\hat g_{3}, \hat g'_{3} \neq 0$ with the one given in \cite{Lu:2018suj} for which $\hat g_{3} = \hat g'_{3} = 0$ (also $\nu_{2} = 0$) to see the effect of the added fluxes $\hat g_{3}, \hat g'_{3} \neq 0$. This latter rate is
\be\label{p3rate}
{\cal W} (\hat g_{3} = \hat g'_{3} =0) \approx \frac{8\pi |\hat f_1 - \hat f'_1|\nu_{1}}{(8 \pi^2 \alpha')^{{\frac{p+1}{2}}}}\,  \nu_0^{\frac{p- 3}{2}}  e^{-\frac{ y^2}{2\pi \alpha' \nu_{0}}}\,  \frac{\left[\cosh\frac{\pi \nu_{1} }{\nu_{0}} +1 \right]^{2}}{\sinh\frac{\pi \nu_{1} }{\nu_{0}}},
\ee
which is valid for $p \ge 3$.  For each given $p \ge 5$, we have
\be\label{rate-ratio}
\frac{{\cal W} (\hat g_{3}, \hat g'_{3} \neq 0)}{{\cal W}  (\hat g_{3} = \hat g'_{3} =0)} = \frac{\pi \nu_{2}}{\nu_{0}} \frac{\left[\cosh\frac{\pi (\nu_{1} + \nu_{2}) }{\nu_{0}} +1 \right]\left[\cosh\frac{\pi (\nu_{1} - \nu_{2})}{\nu_{0}} + 1\right]}{ \left[\cosh\frac{\pi \nu_{1} }{\nu_{0}} +1 \right]^{2}\sinh\frac{\pi \nu_{2} }{\nu_{0}}}.
\ee
Let us examine this ratio. For example, if $\pi \nu_{2}/\nu_{0} \ll 1$ while holding $\pi \nu_{1}/\nu_{0}$ fixed, we have it to be unity.  In other words, adding the additional fluxes doesn't help. We consider now the other case when $\pi \nu_{1}/\nu_{0} \gg 1$ and  $\pi \nu_{2}/\nu_{0} \sim {\cal O} (1)$ or larger. Now the above ratio $\sim (\pi\nu_{2}/\nu_{0}) / \sinh (\pi\nu_{2})/\nu_{0}) < 1$. Further when $\pi\nu_{2}/\nu_{0} \gg 1$ (noting $\nu_{1} \ge \nu_{2}$), this ratio becomes vanishingly small.  So adding magnetic fluxes $\hat g_{3}$ and $\hat g'_{3}$ in general diminishes rather than enhances the rate. 

Let us enforce this further.   As discussed in detail in \cite{Lu:2017tnm} in the absence of $\hat g_{3}$ and $\hat g'_{3}$,  the ratio of dimensionless rates for $p = 5$ and for $p = 3$ for the same applied fluxes is 
\be\label{p5p3r}
r = \frac{ (2 \pi \alpha')^{3} {\cal W}^{p = 5}}{(2\pi \alpha')^{2} {\cal W}^{p = 3}} = \frac{\nu_{0}}{4 \pi},
\ee
which can also be read from the rate given in (\ref{p3rate}). Let us give an estimate how small this ratio is.  For simplicity, we assume $\hat f'_{1} = 0$ and $\hat f_{1} = 2 \pi \alpha' E$ with  $E$ the laboratory electric field.  Note that the current constant electric field limit in laboratory is  8 order of magnitude smaller than the one required for producing the Schwinger pair production which is $E \sim m^{2}_{e} \sim 10^{- 7} {\rm GeV}^{2}$ with $m_{e}$ the electron mass. The  current constraint for string scale $M_{s} = 1/\sqrt{\alpha'}$ is from a few TeV upto the order of $10^{16} \sim 10^{17}$ GeV (for example, see \cite{Berenstein:2014wva}).  If we take one of D3 branes as our own world,  even taking $E \sim 10^{- 7} {\rm GeV}^{2}$ and $\sqrt{\alpha'} = 1/M_{s}  \sim 10^{- 3} \,{\rm GeV}^{-1}$, we have  the above ratio $ r = \nu_{0}/4\pi \sim 10^{- 11}$, very tiny.  Note that adding the additional magnetic fluxes makes us to move from $p = 3$ to $p = 5$. This is worth the effort only if the further enhancement is much greater than $1/r \sim 10^{11}$. However, our earlier discussion above says that this is impossible. In other words, adding fluxes $\hat g_{3}$ and $\hat g'_{3}$, giving $\nu_{2} \le \nu_{1}$, has no advantage at all over the $p = 3$ rate in the absence of them.  

So combining this with the previous studies given in \cite{Lu:2017tnm,Lu:2018suj} in the absence of $\hat g_{3}$ and $\hat g'_{3}$, we must conclude that the largest rate occurs for the system of two D3 branes considered when the applied fluxes remain the same for all $p \ge 3$.  This gives us also the advantage to actually test this pair production since one of the D3 can be our own world. 

\section{Discussion and conclusion}
In this paper,  we consider a system of two Dp branes, placed parallel at a separation and with each carrying certain electric and magnetic fluxes, and  seek to obtain further enhancement of the open string pair production in the aim of its practical use such as its potential detection. Our previous experience indicates that the most efficient and direct way realizing this is to add the additional magnetic fluxes $\hat g_{3}$ and $\hat g'_{3}$ in a way as given in (\ref{F}) and (\ref{F'}).  However, on the contrary, it turns out that adding such magnetic fluxes diminishes rather than further enhances the enhanced pair production in the absence of them.   

This has important implications. It first shows that for each given $p \ge 3$ and the same applied electric fluxes $\hat f_{1}$ and $\hat f'_{1}$, the largest possible open string pair production is the one when we set $\hat g_{3} = \hat g'_{3} = 0$ in (\ref{F}) and (\ref{F'}), and keep the magnetic fluxes $\hat g_{2}$ and $\hat g'_{2}$ to give the largest possible $\nu_{1}$ in practice. Combining this with the previous results of the enhanced open string pair production in the absence of $\hat g_{3}$ and $\hat g'_{3}$, we conclude that the largest pair production is for the $p = 3$ case if we keep the same electric and magnetic fluxes for all allowed $p \ge 3$ cases.  This is interesting since for $p = 3$, one of D3 branes can be our own 4-dimensional world and this  also 
makes the detection of the pair production a potential reality, an advantage over any other $p > 3$ cases. As discussed recently in \cite{Lu:2018nsc},  from the perspective of a brane observer, the open string pair appears as the charged particle/anti-charged particle one and the dimensions transverse to the brane are the extra-dimensions. So a positive detection of this pair production by the brane observer will indicate the existence of extra-dimensions since the open strings so produced, as mentioned earlier, are those connecting the two D3 branes in the directions transverse to the branes.  This will further give a test of the underlying string theories since the computations are stringy ones.

\section*{Acknowledgements}
We would like to thank Zihao Wu, Li Zhao and Xiaoying Zhu for discussions. The authors acknowledge support by grants from the NSF of China with Grant No: 11775212 and 11235010.

\end{document}